\documentclass[twocolumn,english,showpacs,prl]{revtex4}
\usepackage[T1]{fontenc}
\usepackage[latin9]{inputenc}
\usepackage{amsmath}
\usepackage{graphicx}
\usepackage{amssymb}

\makeatletter
\@ifundefined{textcolor}{}
{%
 \definecolor{BLACK}{gray}{0}
 \definecolor{WHITE}{gray}{1}
 \definecolor{RED}{rgb}{1,0,0}
 \definecolor{GREEN}{rgb}{0,1,0}
 \definecolor{BLUE}{rgb}{0,0,1}
 \definecolor{CYAN}{cmyk}{1,0,0,0}
 \definecolor{MAGENTA}{cmyk}{0,1,0,0}
 \definecolor{YELLOW}{cmyk}{0,0,1,0}
 }

\usepackage{dcolumn}\usepackage{bm}\usepackage{amsfonts}\usepackage{times}\usepackage{babel}\usepackage{babel}

\makeatletter
\@ifundefined{textcolor}{}
{
 \definecolor{BLACK}{gray}{0}
 \definecolor{WHITE}{gray}{1}
 \definecolor{RED}{rgb}{1,0,0}
 \definecolor{GREEN}{rgb}{0,1,0}
 \definecolor{BLUE}{rgb}{0,0,1}
 \definecolor{CYAN}{cmyk}{1,0,0,0}
 \definecolor{MAGENTA}{cmyk}{0,1,0,0}
 \definecolor{YELLOW}{cmyk}{0,0,1,0}
 }
\providecommand{\U}[1]{\protect\rule{.1in}{.1in}}
\makeatother

\makeatother

\usepackage{babel}

\begin{document}

\title{The Physical Origins of the Meissner Effect and London Penetration
Depth}

\author{X. Q. Huang$^{1,2}$}

\email{xqhuang@netra.nju.edu.cn}

\affiliation{$^{1}$Department of Physics and National Laboratory of Solid State
Microstructure, Nanjing University, Nanjing 210093, China\\
 $^{2}$Department of Telecommunications Engineering ICE, PLAUST,
Nanjing 210016, China }
\begin{abstract}
Based on the recent developed real-space theory of superconductivity
(arXiv:0910.5511 and arXiv:1001.5067), we study the physical nature
of the Meissner effect and London penetration depth in conventional and non-conventional superconductors. It is argued
that they originate from an exactly the same reason of the real-space quasi-one-dimensional
periodic dynamic charge stripes in the superconductors. The fundamental
relationship between the London penetration depth and the superconducting
electron density is qualitatively determined.
\end{abstract}

\maketitle

\textbf{Introduction:} The Meissner Effect is the most fundamental
and the most dazzling property of materials in the superconducting
state \cite{Meissner1933}. To explain the Meissner effect that no
magnetic field exists in the interior of a superconductor, the London
brothers proposed two equations to describe the microscopic electric
field \textbf{E} and magnetic field \textbf{B} within the superconductor
 \cite{London1935}. The first London equation is in fact the simply
Newton's second law for the superconducting electrons:\begin{equation}
\frac{\partial\mathbf{J_{\mathbf{s}}}}{\partial t}=\frac{n_{s}e^{2}}{m_{e}}\mathbf{E},\label{london1}\end{equation}
 where $\mathbf{J}_{s}=n_{s}e\mathbf{v}_{s}$ is the supercurrent
density, $n_{s}$ is the superconducting electron density and \textbf{$\mathbf{v}_{s}$}
is the superfluid velocity. Taking into account Maxwell's equation
of $\nabla\times\mathbf{E}=-\partial\mathbf{B}/\partial t$, then
the second London equation can be written as \begin{equation}
\nabla\times\mathbf{J_{\mathbf{s}}}=-\frac{n_{s}e^{2}}{m_{e}}\mathbf{B}.\label{london2}\end{equation}

The London equations (\ref{london1}) and (\ref{london2}) imply a
characteristic length scale $\lambda_{L}=\sqrt{m_{e}/\mu_{0}n_{s}e^{2}}$,
the so-called London penetration depth, which gives the length scale
over which an external applied field is exponentially suppressed.
For most superconductors, the London penetration depth $\lambda_{L}$
is on the order of 100 nm. In 1950, Ginzburg and Landau (GL) developed
the order parameter phenomenological theory to describe the superconducting
phenomena in the conventional superconductors \cite{GL1950}. They
derived a temperature dependent penetration depth as $\lambda(T)=\sqrt{m^{*}c^{2}/4\pi e^{2}\left|\Psi\right|^{2}}$,
where $\Psi$ is the superconducting order parameter. While it is
important to note that although these theories may offer part of the
answer to the problems of the Meissner effect and penetration-depth
effect, the exact physical origin of these effects remains a mystery
so far. In addition, the theoretical foundation of the conventional
London equations is not solid enough \cite{Hirsch2004}, as we know,
there is a certain intuitive logic to this approach.

In our opinion, the phenomenological GL theory may include more important
and rigorous information than the later-developed microscopic BCS
theory \cite{BCS1957}. The former was based on a coherent picture
of all charge carriers, while the latter considered only the pairing
behavior of the carriers. Physically, the pairing of the charge carriers
in materials is an individual behavior characterized by pseudogap,
while superconductivity is a collective behavior of many coherent
charge carriers. Obviously, the BCS theory is lack of a convincing
mechanism to ensure the coherence of the Cooper pairs. As more and
more Cooper pairs, the backbone of the superconducting phenomenon,
have now been observed in non-superconducting and insulation materials \cite{Stewart2007}.
These interesting experiments strongly indicate that the BCS is likely
to be physically incomplete.

In the past several years, we have published a series of articles
which offer a new way of looking at the superconducting phenomena
 \cite{Huang1,Huang2,Huang3,Huang4}. It is now very clear that pairing
(pseudogap) is a real space Coulomb confinement effect within single
unit cell \cite{Huang3,Huang4}, while superconducting is related to
the long-range order structures of electrons in real space \cite{Huang1,Huang2}.
Surprisingly, no pairing and superconducting glues are needed in our
scenarios where the direct Coulomb repulsive interactions between
electrons can be naturally and completely suppressed. 

In the present
paper, we try to unveil the mystery of the Meissner effect and London
penetration depth based on our previous works. We will show that the
Meissner effect and London penetration depth are directly originated
from the real-sapce periodic superconducting charge stripes.

\begin{figure}[b]
\begin{centering}
\resizebox{1\columnwidth}{!}{ \includegraphics{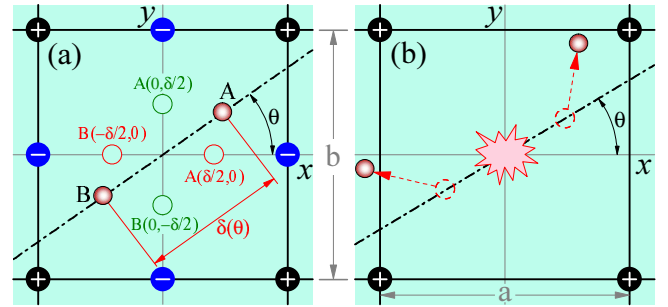}}
\par\end{centering}

\caption{Two electrons (A and B) locate inside a single plaquette, (a)
the hole-doped superconductor, there are four $O^{1-}$ and four $Cu^{2+}$indicated
by solid blue and black circles, respectively, the localized Cooper
pair (pseudogap) may exist inside the plaquette due to the four nearest
neighbor negative ions, (b) the conventional superconductor, where
the localized Cooper pair is impossible to survive in the plaquette
due to the Coulomb attractions by the positive ions. }

\label{fig1}
\end{figure}

\begin{figure*}[t]
\begin{centering}
\resizebox{1.9\columnwidth}{!}{ \includegraphics{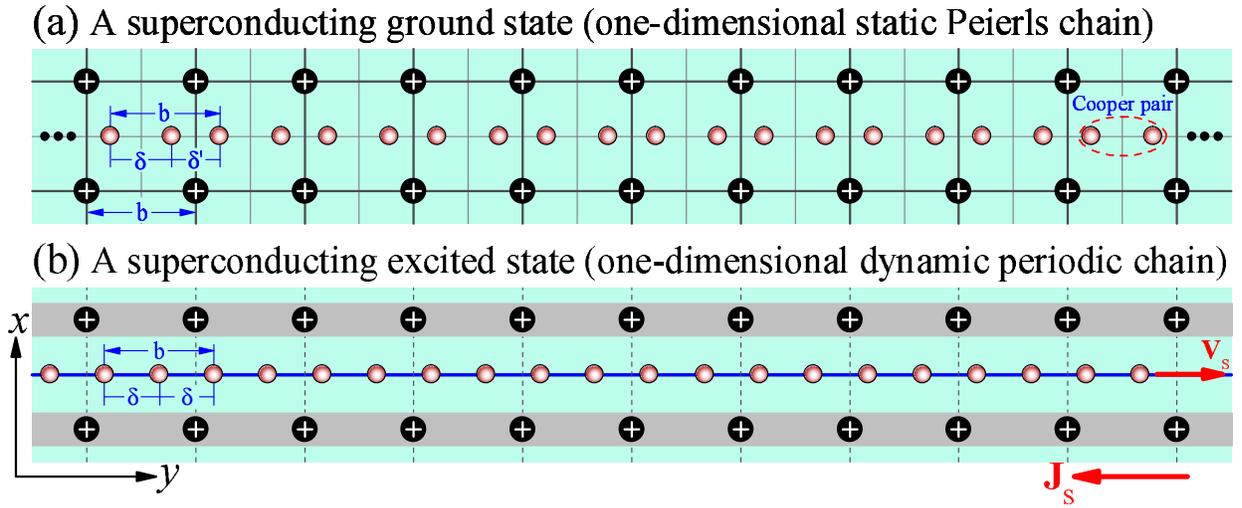}}
\par\end{centering}

\caption{Two kinds of quantized one-dimensional charge stripes. (a). Without
the driven of external fields, the electrons can arrange themselves
into some static one-dimensional Peierls chains with two electrons
(a real-space Cooper pair) inside one plaquette, we have proved analytically
that the repulsive interactions among electrons have been completely
suppressed with the appropriate $\delta$ and $\delta'=b-\delta$.
In the presence of the external fields, there will happen a charge
order phase transition from the static Peierls charge stripe (superconducting
ground state) to the dynamic periodic charge stripe (excited superconducting
state). In this case, the electric field of the discrete positive
ions are replaced by the uniform mean-fields which are regarded as
domain walls and the electron-electron Coulomb repulsions can be naturally
eliminated due to the symmetry of the stripe.}

\label{fig2}
\end{figure*}

\textbf{Real Space Charge Stripes:} Normally, electrons repel each
other according to Coulomb's law, they are attracted only to protons
or positive ions. However, it is argued that strong electron-electron
repulsions should be eliminated in some particular situations like
superconductivity. How could the direct repulsive forces be completely
suppressed in the superconductors?

As shown in Fig. \ref{fig1}, we have a new window on this critical
question of what holds two electrons together \cite{Huang3,Huang4}.
For the hole-doped cuprates, we argued that a localized hole-pair
is a cluster of two electrons (a localized Cooper pair), four $O^{1-}$
and four $Cu^{2+}$ inside the Cu-O plane, as illustrated in Fig.
$\ref{fig1}$(a). It is easy to prove that the direct and strong electron-electron
repulsion can be entirely excluded if two electrons are located symmetrically
at {[}$A(\delta/2,0)$, $B(-\delta/2,0)${]} in $x$-direction or
{[}$A(0,\delta/2)$, $B(0,-\delta/2)${]} in $y$-direction within
each copper-oxide unit with the Cooper-pair size $\delta\approx0.396b$
(when $a=b$), as shown in Fig. $\ref{fig1}$(a). Moreover, the four
nearest-neighbor electron-$O^{1-}$ repulsive interactions play the
key role of the `pairing glue' for the real-space localized Cooper
pair. Hence, it should be noted that the existence of the negative
ions is the most important condition for supporting the pseudogap
phases in the materials (no limited to the superconductors) \cite{Huang1,Huang3}.
This simple and reliable approach leads to the $d$-wave symmetry
pseudogap behavior in the hole-doped cuprates. For the conventional
superconductors of Fig. $\ref{fig1}$(b), the two electrons cannot
maintain pairing in the proposed structure due to the direct electron-ion
Coulomb attractions, as a result, these materials do not support the
pseudogap's electronic state \cite{Huang3,Huang4}.

\begin{figure}[b]
\begin{centering}
\resizebox{1\columnwidth}{!}{ \includegraphics{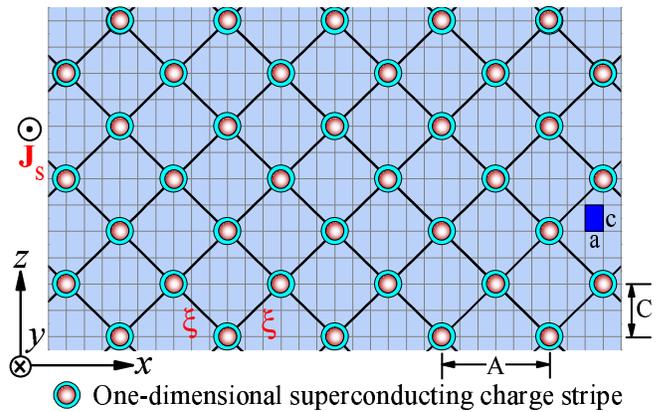}}
\par\end{centering}

\caption{One of the stable structure of the superconducting vortex lattices
(quasi-two-dimensional charge stripes), where the electron-electron
repulsions among different stripes can also be canceled out because
of the symmetry structure.}

\label{fig3}
\end{figure}

It is now widely believed that the electron pairing (Cooper pair)
is not the most fundamental reason for the superconductivity \cite{Stewart2007,Anderson2007,Lawler2010}.
The key to solve the superconducting puzzle lies in the problem of
how can the Coulomb repulsions between pairs of electrons be overcome
in favor of the superconductivity? In fact, the answer is very simple:
symmetry and minimum energy principle. Generally, a superconductor
is composed of two parts: the periodically arranged positive ions
and the electrons (see Fig. $\ref{fig2}$), which only interact via
the electromagnetic interaction. Based on the real space Coulomb confinement
effect, we proved theoretically that the one-dimensional superconducting
ground state (a static Peierls chain with $\mathbf{v}_{s}=0$) and
the excited state (a dynamic periodic charge stripe with the electron-electron
distance $\delta=b/2$ and $\mathbf{v}_{s}\neq0$) can be symmetry
and naturally formed inside one superconducting plane, as shown in
Figs. $\ref{fig2}$(a) and (b) respectively. In our scenarios, all
the superconducting electrons are assumed to be in a zero-stress state
(the electron-electron repulsions are naturally suppressed) and can
be regarded as the `inertial electrons'. The quasi-one-dimensional
charge stripes of Fig. $\ref{fig2}$ may further self-assemble into
some stable quasi-two-dimensional Wigner structures with trigonal
or tetragonal symmetries \cite{Huang1,Huang2}, see Fig. \ref{fig3}
as an example.

\textbf{Meissner Effect and London Penetration Depth:} We now firmly
believe that all superconducting phenomena may come from an identical
physical reason, which we think may be due to  the ordered structures
of the charge carriers in real space (see Fig. $\ref{fig2}$ and Fig.
$\ref{fig3}$). In this section, we try to explain the observed facts
of Meissner effect and London penetration depth in various superconductors
using the suggested model.

\begin{figure}
\begin{centering}
\resizebox{1\columnwidth}{!}{ \includegraphics{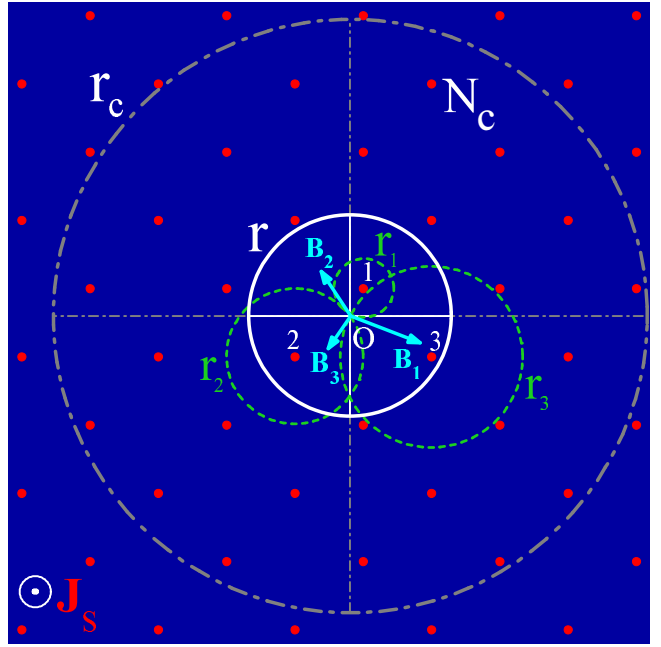}}
\par\end{centering}

\caption{The explanation of the Meissner Effect.\textbf{ }In our framework,
the central feature in the superconductor is the real-space infinite
charge stripes (indicated by red solid circles). Each stripe will
generate a magnetic field at any location inside the superconductor
(for example the `O' point in the figure). Approximately, the net
magnetic field of `O' can be considered as the contribution from the
limited number of charge stripes inside a critical circle $r_{c}$.
As a result, the total magnetic field at point `O' is expected to
be zero due to the symmetry. }

\label{fig4}
\end{figure}

In our framework, the so-called superconducting electron states are
in fact the periodic arrays of aligned infinite straight wires (charge
stripes). Suppose there are $N$ dynamic charge stripes (each with
a current $I_{0}$) distributed evenly in a superconductor, then the
magnetic field at any location (indicated `O' in the Fig. \ref{fig4})
within the material can be expressed as:\begin{equation}
\mathbf{B}_{O}=\sum_{i=1}^{N}\mathbf{B}_{i},\end{equation}
where the value $B_{i}=\mu_{0}I_{0}/2\pi r_{i}$, obviously, the magnetic
field $B_{i}$ will decrease rapidly with the increasing of $r_{i}$.

Strictly speaking, the magnetic field at any point should take into
account all contributions of the charge stripes in the superconductor.
But due to the rapid attenuation of the magnetic field, only a limited
number of the charge stripes around the observation point will contribute
to the magnetic field at point `O' of Fig. \ref{fig4}. For a small
radius $r$ (see Fig. \ref{fig4}), only a few charge stripes are
included and the total magnetic field generated by them is usually
not equal to zero. As the radius $r$ increases, more charge stripes
produce magnetic fields at the circle center `O', however, the total
magnetic field will decay rapidly because of the symmetry of the intensive
charge stripes inside the circle. This implies that there exists a
critical radius $r_{c}$ (or a critical charge stripes number $N_{c}$
for a given supercurrent density $\mathbf{J}_{s}$) that will make
the total magnetic field $\mathbf{B}_{o}=\sum_{i=1}^{N_{c}}\mathbf{B}_{i}\approx0$,
naturally lead to the Meissner effect.

In the following, we will turn to a simple discussion of the London
penetration depth. By relating the second London equation (\ref{london2})
to Maxwell's equations, and we obtain\begin{equation}
\nabla^{2}\mathbf{B}=\frac{\mu_{0}n_{s}e^{2}}{m_{e}}\mathbf{B}=\frac{1}{\lambda_{L}^{2}}\mathbf{B}.\label{depth}\end{equation}

For the simple one-dimensional ($x$) geometry, then we can obtain
the solution from Eq. (\ref{depth}), that is $B(x)=B_{0}e^{-x/\lambda_{L}}$,
or $B(\lambda_{L})=B_{0}/e$ indicates that the magnetic field at
the surface of the superconductor will decay to $B_{0}/e$ at a distance
$\lambda_{L}$ in the interior of the superconductor. Therefore, the
London equations indeed lead to an exponential decay of the magnetic
field within the superconductor. In our view, although the London
equations provide a phenomenological description of the electromagnetic
properties inside the superconductor, the most essential reason of
this phenomenon is still unknown. This article demonstrates that the
correct answer still lies in the nature of the charge orders of Fig.
\ref{fig2} and Fig. \ref{fig3} in real space.

\begin{figure}
\begin{centering}
\resizebox{1\columnwidth}{!}{ \includegraphics{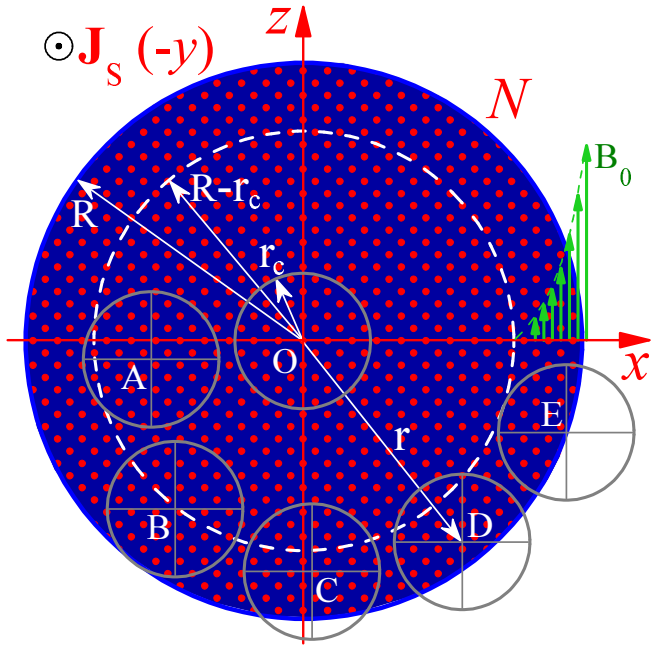}}
\par\end{centering}

\caption{The explanation of the London penetration depth (see the text for
details). }

\label{fig5}
\end{figure}

We assume that the shape of the studied superconductor is a cylinder
with a cross section radius $R$ and a superconducting current $\mathbf{J}_{s}$
in $-y$ direction, as shown in Fig. \ref{fig5}. As discussed above,
there is an area ($r\leq R-r_{c}$, where $r_{c}$ is the critical
radius) inside the superconductor where the magnetic field is almost
zero, as marked by the white dotted circle in the figure. When $r>R-r_{c}$,
the critical circles no longer fall entirely within the superconductor
(see the cases of C, D and E in Fig. \ref{fig5}), in these situations,
the magnetic fields of the corresponding circle centers will not equal
to zero. With $r$ increasing from $R-r_{c}$ to $R$, the net magnetic
field will increase simultaneously from zero (the case B in Fig. \ref{fig5})
to the maximum value $B_{0}$ (the case E in Fig. \ref{fig5}). Hence,
the critical radius $r_{c}$ can be regarded as the London penetration
depth is associated with the superconducting electron density $n_{s}$.
In our model, for a given superconductor, the structure of each superconducting
line {[}see Fig. \ref{fig2} (b){]} is independent of $n_{s}$, thus
the number of the charge stripes $N$ is proportional to $n_{s}$.
Our discussions above are based on the assumption that the critical
circle ($r{}_{c}$) should cover approximately $N_{c}$ charge stripes,
and hence we have the following expression: \begin{equation}
\pi r_{c}^{2}n_{s}=\beta N_{c}=Constant,\label{ns}\end{equation}
where $\beta$ is a constant. If we replace $r_{c}$ with the penetration
depth $\lambda_{L}$ in Eq. (\ref{ns}), this immediately lead to
the relationship $\lambda_{L}\propto1/\sqrt{n_{s}}$ which is in agreement
with that of London equations. Notice that $r_{c}$ is not equal to
$\lambda_{L}$, since the critical parameter $r_{c}$ is defined as
$B(r_{c})/B_{0}\rightarrow0$ different from $B(\lambda_{L})/B_{0}=1/e\approx37\%$
of the $\lambda_{L}$. Moreover, we think that the parameter $\lambda_{L}$
does not describe accurately enough the magnetic flux penetration
behavior in the superconductors. The real penetration depth is much
larger than that determined by the London equations. Hence we introduce
a new parameter $\lambda(n)$ which is defined by $\lambda(n)=n\lambda_{L}$
($n=$1, 2, 3, 4...), then the attenuation of the magnetic field can
be expressed as $B[\lambda(n)]/B_{0}=1/e^{n}$. When $n=3$, then
$\lambda(3)=3\lambda_{L}$ and $B[\lambda(3)]/B_{0}=1/e^{3}\approx5\%$
which is relatively more suitable for describing the experimental
results.

It should be emphasized that our interpretations about the Meissner
effect and London penetration depth\textbf{ }are primarily based on
the simple and ideal picture of the charge stripes described in Fig.
\ref{fig2} and Fig. \ref{fig3}, many factors such as temperature,
crystal structures, the size and shape of the materials are not taken
into consideration. In our theory the critical charge stripe number
$N_{c}$ is a very important parameter which can be determined approximately
by the experimental results. For the charge stripe structure of Fig.
\ref{fig3}, each charge stripe occupies a area of $\xi^{2}$ in $xz$-plane.
According to our previous paper of the effective $c$-axis lattice
constant  theory of superconductivity  \cite{Huang2}, the stripe-stripe spacing $\xi$
is closely related to the  $c$-axis lattice constant of the superconductors.
For conventional superconductors such as Pb of the cubic close-packed
(CCP) structure with the lattice constants $a=b=c=0.495$ nm, it is
likely that the charge carriers (electrons) form the superconducting
charge stripe pattern of Fig. \ref{fig3} with the minimum value of
$\xi^{Pb}=c/\sqrt{2}\approx0.35$ nm. In addition, the experimental
value of London penetration depth $\lambda_{L}$ for Pb is about 40
nm, approximatively, we have the critical radius $r_{c}^{Pb}\approx\lambda(3)=3\lambda_{L}=120$
nm. Then the critical number $N_{c}$ for Pb can be estimated as $N_{c}=\pi\left(r_{c}^{Pb}/\xi^{Pb}\right)^{2}\approx369000$.
One can find this critical $N_{c}$ is a fairly large value, implying
that the Meissner effect does not exist in the ultra-small superconductor
grains. For the cuprate superconductors such as $Hg_{2}Ba_{2}Ca_{3}Cu_{4}O_{10}$,
the stripe-stripe spacing $\xi$ can reach about $1.9$ nm which
is about five times that of the conventional superconductors. As a
result, the cuprate superconductors normally have a larger London
penetration depth than the conventional superconductors.

\textbf{Brief Summary:} In this paper, the physical origins of the
Meissner effect and London penetration depth observed in the superconducting
materials have been studied. In our new framework, the Meissner effect
and London penetration depth are the straightforward results of the
real-space ordered charge stripes within the superconductors. We believe
that the so-called complex phenomenon of superconductivity can be
well understood from the simplest models and the most basic physical
theories. Any artificial quasiparticles are in fact obstruct our understanding of phenomenon of superconductivity.

\end{document}